\documentclass[titlepage,12pt]{utarticle}
\usepackage{amssymb}
\numberwithin{equation}{section}

\newcommand{\abs}[1]{\lvert #1\rvert}

\newcommand{\BW}{\field{W}}

\begin{document}
\preprint{
  UTTG--06--03\\
 }

\title{Finding  the Mirror of the Beauville Manifold}
\author{
 Hyukjae Park%
    \thanks{Work supported in part by NSF Grant PHY0071512}
}
 \oneaddress{
      Theory Group, Physics Department\\
      University of Texas at Austin\\
      Austin, TX 78712 USA\\{~}\\
      \email{hpark@zippy.ph.utexas.edu}
}

\date{December 5, 2003}

\Abstract{
We construct the mirror of the Beauville manifold. The Beauville manifold is a Calabi-Yau manifold with non-abelian fundamental group. We use the conjecture of Batyrev and Borisov to find the previously misidentified mirror of its universal covering space, $\BP^7[2,2,2,2]$. The monomial-divisor mirror map is essential in identifying how the fundamental group of the Beauville manifold acts on the mirror of $\BP^7[2,2,2,2]$. Once we find the mirror of the Beauville manifold, we confirm the existence of the threshold bound state around the conifold point, which was originally conjectured in   \cite{Brunner:2001sk}. We also consider how the quantum symmetry group acts on the D-branes that become massless at the conifold point and show the action proposed in \cite{Brunner:2001eg} is compatible with mirror symmetry.}

\maketitle


\section{Introduction}
In the last one and a half decades or so, there has been great progress in understanding the non-perturbative 
aspects of string theory. Although string theory, as it exists now, lacks a complete
definition, one can still study various non-perturbative effects through tools like supersymmetry and dualities.
Mirror symmetry is one of these dualities. Mirror symmetry is a duality between Type IIA string theory compactified on a Calabi-Yau manifold and Type IIB on another Calabi-Yau, called the first manifold's mirror   \cite{Morrison:1996yi}. Since it was first proposed in   \cite{Greene:1990ud}, it has been extensively studied. Many examples of Calabi-Yau manifolds have been studied and their mirrors have been found   \cite{Berglund:1994ax, Candelas:1994dm, Candelas:1994hw}. 
However, all examples studied so far have abelian fundamental groups. Indeed, there is only one known example of  Calabi-Yau manifolds with $SU(3)$ holonomy(and not a proper subgroup thereof) whose fundamental groups are non-abelian. That is the manifold Beauville constructed in   \cite{Beauville:1995}. 

There are several reasons that Calabi-Yau  manifolds with non-abelian fundamental groups are interesting.
These kinds of Calabi-Yau manifolds may be used to construct  phenomenologically realistic models. 
When heterotic string theory is compactified on a Calabi-Yau manifold, the gauge group is broken down to the subgroup that is preserved by the vacuum expectation values of the gauge fields. The vacuum expectation values should be chosen carefully to preserve supersymmetry and to satisfy the anomaly cancelation condition. Most of the times, the gauge group so obtained is too big to be phenomenologically interesting. One can break the gauge group further by turning on Wilson lines. However, if the fundamental group of the Calabi-Yau manifold is abelian, turning on Wilson lines does not change the rank of the gauge group. Hence, if we want to reduce the rank of the gauge group, a non-abelian fundamental group is necessary   \cite{Green:1987mn}. 

Also, Calabi-Yau manifolds with non-abelian fundamental groups provide good examples for studying various 
aspects of D-branes. For example, in   \cite{Brunner:2001sk}, it is conjectured that, on the level of K-theory, the monodromy about the conifold locus (principal component of the discriminant locus) is  of the form:
\begin{equation}
 v \to v-\sum_R (v, W_R)W_R
\end{equation}
where the sum is over all irreducible representations, $R$, of the fundamental group, $G$, and $W_R$ is the flat bundle built using the irreducible representation, $R$.
If $G$ is non-abelian, it implies there are threshold bound states of D6-branes that are stable and become massless at the conifold locus. Since the existence of these states are guaranteed by neither the BPS condition nor by K-theory, it is interesting to study these states. 

Another interesting application is the action of the quantum symmetry group on D-branes. Any non-simply connected Calabi-Yau manifold, $X$ can be written as $X = Y/G$. $Y$ is the universal covering space and $G$ is the fundamental group of $X$. Hence, string theory on $X$ is described by an orbifold conformal field theory and this orbifold CFT has a quantum symmetry group, $G/[G,G]$. The action of this group on states in the closed string sector is known and well understood.  It is interesting to see how it acts on non-perturbative states like D-branes. Its action on D-brane's charges in K-theory has been conjectured in    \cite{Brunner:2001eg}. Since it applies to both A-branes and B-branes,  it is  interesting to check the conjectured action is compatible with mirror symmetry.  If the fundamental group is non-abelian, it becomes more interesting since the quantum group is not any longer isomorphic to the fundamental group. 

The Beauville manifold has the fundamental group $Q$, the group of unit quaternions and its universal covering space is $\BP^7[2,2,2,2]$, the intersection of 4 quadrics in $\BP^7$. We will construct the mirror of the Beauville manifold by taking the quotient of the mirror of $\BP^7[2,2,2,2]$ by $Q$. The mirror of $\BP^7[2,2,2,2]$ has been considered in the literature  \cite{Berglund:1994ax}. However, as we will explain later, the conjectured mirror has a significant flaw. In this paper, we construct its correct mirror using the Batyrev and Borisov conjecture  \cite{Borisov:1993,Batyrev:1997tv}. The monomial-divisor mirror map  \cite{Aspinwall:1993rj} is the essential tool in  finding the $Q$-action on the mirror. To do so, we extend the original argument a little bit since $\BP^7[2,2,2,2]$ is a complete intersection Calabi-Yau manifold. 

The procedure we use here  can be applied to any non-simply connected Calabi-Yau manifold $X$. One finds the mirror of the universal covering space $Y$,  and applies the monomial-divisor mirror map to find how the fundamental group $G$, acts on the mirror. There is a subtlety though. The monomial-divisor mirror map is  valid only in the large radius limit. There is no guarantee that $Y$'s mirror is in the neighborhood of the  large radius limit. Also, the definition of monomials depends on the choice of homogeneous coordinates.  Different choices of homogeneous coordinates will give different monomial-divisor mirror maps. Therefore, we need to choose the homogeneous coordinate carefully. 
 
\section{The Beauville Manifold}\label{Beauville}
There is only one known example of Calabi-Yau manifolds with non-abelian fundamental groups. 
It is the manifold Beauville constructed in    \cite{Beauville:1995}. The fundamental group of this Calabi-Yau manifold is  the group of  unit quaternions:
\begin{equation}
  Q=\left\{\pm1,\pm I, \pm J,\pm K\right\}
\end{equation}
with multiplication law
\begin{equation}
\begin{split}
   I J &= K\qquad \text{(and cyclic)}\\
   I^2 &=J^2=K^2=-1
\end{split}
\end{equation}
Before we review the construction of this manifold, let us recall some facts about the group theory of $Q$  \cite{Brunner:2001sk}. First, there is an exact sequence,
\begin{equation}
   0\to \BZ_2\to Q \to \BZ_2\times\BZ_2\to 0
\end{equation}
where the commutator subgroup of $Q$ is the $\BZ_2$ subgroup, $\left\{1,-1\right\}$
and its
abelianization, $Q/[Q,Q]=\BZ_2\times\BZ_2$.

The irreducible representations of $Q$ are as follows. There are four
1-dimensional irreps: the trivial rep $V_1$ and the representations $V_I,V_J,$
and $V_K$. In $V_I$, $\pm1$ and $\pm I$ are represented by $1$ while $\pm J$ and
$\pm K$ are represented by $-1$ (and similarly for $V_{J,K}$).  There is also a
2-dimensional representation, $V_2$. $\pm I, \pm J$ and $\pm K$ act on $V_2$ by 
$\pm i \sigma_3, \pm i \sigma_2$ and $\pm i \sigma_1$.
The representation ring is
\begin{equation}\label{IrrepsRing}
\begin{split}
   V_2\otimes V_2&= V_1\oplus V_I\oplus V_J \oplus V_K\\
   V_\alpha\otimes V_2 &= V_2\qquad \alpha=1,I,J,K\\
   V_I\otimes V_J &= V_K \qquad \text{(and cyclic)}
\end{split}
\end{equation}
The group homology of $Q$ is
\begin{equation}
   \homo{1}{Q}=Q/[Q,Q]=\BZ_2\oplus\BZ_2,\qquad \homo{2}{Q}=0
\end{equation}

Now, let's construct the Beauville manifold. Let $V_8$ be the regular representation of $Q$ and $\BP[V_8]$ be its projective space. The $Q$-action on $V_8$ induces a $Q$-action on $\BP[V_8]$ and 
also on  the space of quadrics in it. We choose 4 quadrics, one from each 1-dimensional irreducible representation. Then, the intersection $Y$, of these quadrics will be invariant under $Q$. 
What Beauville showed in   \cite{Beauville:1995} is that for generic enough choices, $Y$ is a smooth Calabi-Yau manifold and $Q$ acts freely on $Y$. Hence, 
we can take the quotient $X = Y/ Q$ and $X$ is a smooth Calabi-Yau manifold with fundamental group $Q$. 

The Hodge numbers of $Y$ and $X$ are $h^{1,1}(Y)=1$,
$h^{2,1}(Y)=65$ and $h^{1,1}(X) = 1$, $h^{2,1}(X) = 9$, respectively. It is easy to show that $Q$ acts trivially on $\homo{2}{Y}$. Applying the Cartan-Leray Spectral Sequence as in   \cite{Brunner:2001eg}, we can show that
\begin{equation}
\pi_* : \homo{2}{Y} \to \homo{2}{X} 
\end{equation}
is an isomorphism. Here, $\pi_*$ is the push-forward of the projection $\pi : Y \to X$.
The Poincar\'e duality implies that the pull-back
\begin{equation}
\pi^* :\coho{2}{X} \to \coho{2}{Y}
\end{equation}
is also an isomorphism. The classical consideration around the large radius limit shows that the K\"ahler moduli space of $X$ is  the same as that of $Y$  \cite{Aspinwall:1994uj}.  
\section{Mirror of $Y$}
In the literature  \cite{Berglund:1994ax}, the mirror of $Y = \BP^7[2,2,2,2]$ has been conjectured to be $Z = \BP^7[2,2,2,2] / G$ where $G \simeq (\BZ_4)^3$ is generated by\footnote{Actually, in   \cite{Berglund:1994ax}, the group $G$ was not explicitly  given.}
\begin{equation}
\begin{split}
g_1 : [X_1,X_2,X_3,X_4,X_5,X_6,X_7,X_8] & \mapsto [ \zeta X_1, \zeta^3 X_2, \zeta^2 X_3, X_4, X_5,X_6,X_7,X_8]\\ 
g_2 : [X_1,X_2,X_3,X_4,X_5,X_6,X_7,X_8] & \mapsto [ X_1, X_2, \zeta X_3,\zeta^3 X_4, \zeta^2 X_5, X_6,X_7,X_8]\\ 
g_3 : [X_1,X_2,X_3,X_4,X_5,X_6,X_7,X_8] & \mapsto [ X_1, X_2,  X_3, X_4, \zeta X_5,\zeta^3 X_6,\zeta^2 X_7,X_8]
\end{split}
\end{equation}
and $\zeta$ is a fourth root of unity.
To make $Z$  a Calabi-Yau manifold, we must choose 4 quadrics $G_1, \ldots, G_4$ such that they transform under $G$ in the following way.
\begin{equation}
\begin{split}
g_1: [G_1, G_2, G_3, G_4] & \mapsto [\zeta^2 G_1, G_2, G_3, G_4]\\
g_2: [G_1, G_2, G_3, G_4] & \mapsto [ G_1, \zeta^2G_2, G_3, G_4]\\
g_3: [G_1, G_2, G_3, G_4] & \mapsto [G_1, G_2, \zeta^2 G_3, G_4]
\end{split}
\end{equation}
The most general such quadrics are (up to isomorphism):
\begin{equation} \label{EqsZ}
\begin{split}
G_1 &= X_1^2 + X_2^2 - 2 \psi X_3 X_4\\
G_2 & = X_3^2 + X_4^2 - 2 \psi X_5 X_6\\
G_3 & = X_5^2 + X_6^2 - 2 \psi X_7 X_8\\
G_4 & = X_7^2 + X_8^2 - 2 \psi X_1 X_2.
\end{split}
\end{equation}

$Z$ is singular.  There are fixed points since $G$-action on $Z$ is not free. Those singularities are expected since $Z$ must have $h^{1,1} = 65$. However, there are worse singularities. For any value of $\psi$, $Z$ contains points where the transversality of $G_1, \ldots, G_4$ fails.
For example, at $[0,0,0,0,0,\sqrt{\psi}(1+ i),i,1]$, $dG_1 = G_1 = G_2 = G_3 = G_4 = 0$.  Since all those points are also fixed by some elements of $G$, one might try to resolve these singularities by blowing up.\footnote{Mathematically, blowing up is a procedure for replacing a point with $\BP^{n-1}$ where $n$ is the dimension of the variety we are considering. In this paper, we generalize the notion to include  replacing a subvariety of codimension more than 2 with a subvariety of codimension one which is not necessarily $\BP^{n-1}$. }  Unfortunately, it does not work. Some orbifold singularities can be cured by blow-ups without changing the canonical class  \cite{Greene:1997}. But, in our case, blowing up the singular locus will change the canonical class. That means that the resulting manifold is not any more a Calabi-Yau. Later, we will see how $Z$ is related to the actual mirror. 

To construct the mirror, we use the conjecture originally made by Borisov  \cite{Borisov:1993} and further developed by Batyrev and Borisov  \cite{Batyrev:1997tv}. From now on, we will use toric geometry heavily. For the notation and review, see, for example,   \cite{Fulton:1993, Berglund:1994qk,Greene:1997} and references therein. 

 \subsection{Toric description of $Y$}
Now, we describe $\BP^7$ as a toric variety. 
Let $N$ be a lattice of rank 7 and $e_1, \ldots, e_7$ be its generators. We define $e_8 = -e_1-\ldots-e_7$ and also denote by $N_\BR$ the real scalar extension of $N$. Let  $\Sigma$ be the fan in $N_\BR$  whose cones are generated by proper subsets of the vectors $e_1, \ldots, e_8$. The toric variety associated to this fan is $\BP^7$. Another way of realizing $\BP^7$ as a toric variety is using a polyhedron. Consider the polyhedron $\Delta^* = \mathrm{Conv}\left(\left\{e_1, \ldots, e_8\right\}\right)$ and take its dual $\Delta$ in $M_\BR$. Here, $M_\BR$ is the real scalar extension of the dual lattice $M$. Then, $\BP_{\Delta} = \BP^7$.

To represent $Y$, the intersection of 4 quadrics, $G_1, \ldots, G_4$, we choose the following nef partition of vertices of $\Delta^*$:
\begin{equation}
E_l = \left\{e_{2l-1}, e_{2l}\right\}, \qquad l = 1, \ldots, 4
\end{equation}
Choosing the nef partition amounts to specifying one monomial for each equation, $G_l$. To see this, we need some facts about divisors in a toric variety (see \cite{Fulton:1993}, section 3.3 for fuller explanation). First, recall that every toric variety has the torus action $T = N \otimes_\BZ \BC^*$. 
In study of toric varieties, we are mainly interested in $T$-stable divisors. For Weil divisors, 
we need codimension one irreducible subvarieties that are invariant under $T$-action. There are only finite number of such subvarieties and they are represented by one-dimensional cones, or edges, in the fan. Call them $D_i$. Then, $T$-stable Weil divisors are sums $\sum n_i D_i$ for integers $n_i$. Cartier divisors we are interested in are the ones that are mapped to themselves up to multiplicative constants under $T$ so that their zeroes and poles are invariant. In the toric variety, they are described by continuous integral $\Sigma$-piecewise linear functions. Such a function $\psi$ is defined by specifying $\psi|_\sigma \in M$ for each full dimensional cone $\sigma$ in $\Sigma$. Of course, $\{\psi|_\sigma\}$ must satisfy the continuity condition:
\begin{equation}
\left(\psi|_{\sigma_1} - \psi|_{\sigma_2}\right) \perp \sigma_1 \cap \sigma_2
\end{equation}
 Note that for every $m \in M$, there is a corresponding meromorphic function $\chi^m$.
Using this fact, we construct a Cartier divisor from $\{\psi|_\sigma\}$. Open sets are ones given by the full dimensional cones $\sigma$ in the fan and local equations are $\chi^{-\psi|_\sigma}$. (Here, the $-$ sign is the convention widely used in the  literature.) A Cartier divisor determines a Weil divisor. In our case, a Cartier divisor given by $\psi$ will give  Weil divisor $\sum - \psi(v_i) D_i$ where $v_i$ is the first lattice point met along the edge representing $D_i$. Conversely, for a given Weil divisor $\sum n_i D_i$, we get a Cartier divisor if the toric variety is smooth. The representing  function $\psi$ of the Cartier divisor is uniquely determined by the condition $\psi(v_i) = -n_i$. A Cartier divisor $D$ also determines a lattice convex polyhedron in $M_\BR$ defined by
\begin{equation}
\Delta_D = \left\{y \in M_\BR\; | \; \langle x, y \rangle \ge \psi(x) \quad\forall x \in N_\BR \right\}
\end{equation}
This polyhedron is called the support of global sections of $\sheaf{O}(D)$ since they (the global sections) are determined by the lattice points inside $\Delta_D$:
\begin{equation}
\Gamma\left(X, \sheaf{O}(D)\right) = \bigoplus_{m \in \Delta_D \cap M} \BC \cdot \chi^m.
\end{equation}
 
 Now, let's apply the above theory to our case. In $\Sigma$, there are 8 edges, $\tau_i$, $i = 1, \ldots, 8$. Each $\tau_i$ is  generated by vertex $e_i$ of $\Delta^*$ and represents divisor $D_i = \{X_i = 0\}$. The nef partition above determines 4 Weil divisors:
\begin{equation}
W_l = D_{2l-1} + D_{2l}
\end{equation}
These Weil divisors, in turn, determine Cartier divisors. Their representing functions $\psi_1, \ldots, \psi_4$ are given by
\begin{equation}\label{Psi}
\left.\psi_l \right|_{\sigma_i} = 2 \omega_i - \omega_{2l-1} - \omega_{2l}.
\end{equation} 
Here, $\sigma_i$, $i = 1, \ldots, 8$ are full dimensional cones in $\Sigma$. Each $\sigma_i$ is generated by $e_1, \ldots, \makebox[0pt][l]{\,/}e_i, \ldots, e_8$. Also, $\{\omega_1, \ldots, \omega_7\}$ is the dual basis of $\{e_1, \ldots, e_7\}$ and $\omega_8 = 0$. The partition being nef implies $\psi_l$'s are convex functions. In terms of homogeneous coordinate, these divisors are:
\begin{equation}
X_{2l-1} X_{2l}
\end{equation}
Therefore, $E_l$ represents the monomial $X_{2l-1} X_{2l}$ in $G_l$. 

From the nef partition, we define two sets of  convex lattice polyhedra, $\Pi = \left\{\Delta_1, \ldots, \Delta_4\right\}$, $\Pi^* = \left\{\nabla_1, \ldots, \nabla_4\right\}$.
\begin{equation}
\begin{split}
\Delta_l & = \left\{x \in M_\BR\; |\; \langle x, y \rangle \ge \psi_l(y)\right\}\\
\nabla_l & = \mathrm{Conv}\left(\left\{0\right\} \cup E_l\right) 
\end{split}
\end{equation}
$\Delta_l$ is the support of global sections of $\sheaf{O}_{\BP^7}(W_l)$.
Using the convexity of $\psi_l$, one can show that
\begin{equation}
\Delta_l = \mathrm{Conv}\left(\left\{{\left. \psi_l \right|}_{\sigma_i}\right\}_{ i = 1, \ldots, 8}\right).
\end{equation}
Each $\Delta_l$ contains 36 lattice points:
\begin{equation}
\omega_i + \omega_j -\omega_{2l-1} -\omega_{2l}, \qquad i, j = 1, \ldots, 8
\end{equation}
representing the monomial $X_i X_j$ in equation $G_l$.\footnote{One way of seeing this is to consider the corresponding Weil divisor: 
\begin{equation*}
\sum_{k = 1}^8 (-\psi_l + \omega_i + \omega_j - \omega_{2l-1}-\omega_{2l})(e_k)D_k = D_i + D_j
\end{equation*}
}
Also, note that $0 \in \Delta_l$ and it represents the original monomial $X_{2l-1} X_{2l}$ specified by $E_l$. It is the only common point of $\Delta_1, \ldots, \Delta_4$.

\subsection{Construction of mirror $\widehat{Y}$}
We define a lattice polyhedron $\nabla^* = \mathrm{Conv}\left(\Delta_1 \cup \ldots \cup \Delta_4\right)$ and take its dual $\nabla$.  Let $\widehat{V}$ be the toric variety $\BP_\nabla$.
One can also describe $\widehat{V}$ using the fan $\widehat{\Sigma} = \mathcal{N}(\nabla)$, the normal fan of $\nabla$.  It is the union of the zero-dimensional cone $\{0\}$ together with the set of all cones
\begin{equation}
\sigma[\theta] = \BR^{\ge 0} \cdot \theta 
\end{equation}
that are supporting the faces $\theta$ of $\nabla^*$. $\widehat{\Sigma}$ has 32 edges, $\widehat{\tau}_{l,i}$, $l = 1, \ldots, 4, i = 1, \ldots, 8$, each generated by the vertex $\left. \psi_l \right|_{\sigma_i}$ of $\nabla^*$.
In   \cite{Borisov:1993}, Borisov showed that
\begin{equation}
\widehat{E}_l = \left\{ \left. \psi_l \right|_{\sigma_i}\right\}_{i = 1, \ldots, 8}
\end{equation}
is a nef partition of vertices of $\nabla^*$.
Let  $\widehat{D}_{l,i}$ be the Weil divisor $\widehat{\tau}_{l,i}$ represents. 
As before, each $\widehat{E}_l$ induces a Weil divisor:
\begin{equation}
\widehat{W}_l = \sum_{i = 1}^8 \widehat{D}_{l,i}.
\end{equation}
We choose 4 global sections, $\widehat{G}_l$, one from each $\sheaf{O}_{\widehat{V}}(\widehat{W}_l)$.
Then, the mirror $\widehat{Y}$  of $Y$ is a complete intersection of $\widehat{G}_1, \ldots, \widehat{G}_4$. 
Borisov also showed that the supporting polyhedron of global sections of $\sheaf{O}_{\widehat{V}}(\widehat{W}_l)$ is $\nabla_l$. Lattice points in $\nabla_l$ will generate the global sections. Under the mirror symmetry, the roles of $\Pi$ and $\Pi^*$ are interchanged.

\subsection{Geometry of $\widehat{Y}$}
To understand the geometry of the mirror $\widehat{Y}$ better, we describe $\widehat{V}$ as a holomorphic quotient  \cite{Cox:1993fz}.
First, we introduce ``homogeneous coordinates" $\widehat{X}_{l,i}$, $l = 1,\ldots, 4$ , $i = 1, \ldots, 8$. Each homogeneous coordinate $\widehat{X}_{l,i}$ is paired  to a  vertex $\left.\psi_l\right|_{\sigma_i}$ of $\nabla^*$. Now, $\widehat{V}$ can be written as\footnote{Strictly speaking, it only makes sense as a categorical quotient since $\widehat{\Sigma}$ is not simplicial  \cite{Cox:1993fz}. Here, we are actually considering the family of Calabi-Yau manifolds  and $\widehat{V}$ is a special point on the K\"ahler moduli space. For generic points on the moduli space, the quotient above makes sense as a geometric one. Even at those points corresponding to non-simplicial fans,  the mirror would be given by the geometric quotient rather than one given by toric  geometry since it will be realized via a gauge linear sigma model.}
\begin{equation}
\widehat{V} \simeq \frac{\BC^{32} - F_{\widehat{\Sigma}}}{G}
\end{equation}
The above expression needs some explanation. $\BC^{32}$ is the space whose (plain) coordinate functions are $\widehat{X}_{l,i}$. $F_{\widehat{\Sigma}}$ is the subset of $\BC^{32}$ determined by the fan. If we realize this via a gauge linear sigma model, then $F_{\widehat{\Sigma}}$ is the set of excluded points by the $D$-term conditions.
The group $G$ acts on $\BC^{32}$ as follows:
\begin{equation}
\widehat{X}_{l,i} \mapsto \Lambda_{l,i} \widehat{X}_{l,i} \quad\text{(no sum)}
\end{equation}
where $\Lambda_{l,i}$'s are non-zero complex numbers satisfying:
\begin{equation}\label{GCond}
\prod_{l,i} \Lambda_{l,i}^{\langle n, \left.\psi_l\right|_{\sigma_i}\rangle} = 1
\end{equation}
for any $n \in N$.
This group is the complexified gauge group of the gauge linear sigma model. Its phase part is the actual
gauge group and the magnitude part is fixed by $D$-term  \cite{Witten:1993yc}.
Plugging (\ref{Psi}), we get 7 independent equations:
\begin{equation}
\begin{split}
\prod_{j = 1}^7 \Lambda_{m, j}& = \prod_{l = 1}^4 \Lambda_{l, 2m-1}^2 \\
\prod_{j = 1}^7\Lambda_{m, j} & = \prod_{l = 1}^4 \Lambda_{l, 2m}^2
\end{split}
\end{equation}
for $m = 1, \ldots, 4$. One of the above equations follows from the rest. It turns out that
$G$ is isomorphic to $(\BC^*)^{25} \otimes (\BZ_2)^3$. 
The discrete subgroup,  $G_{torsion} \simeq (\BZ_2)^3$, of $G$ non-trivially acts on the homogeneous coordinates. Because of this, $\widehat{X}_{l,i}$'s are not global sections. Only monomials that are invariant under $G_{torsion}$ will be global sections or Cartier divisors. However, 
the zeroes of $\widehat{X}_{l,i}$ is a globally well defined set. It is shown in   \cite{Cox:1993fz}
\begin{equation}
\widehat{D}_{l,i} = \left\{ \widehat{X}_{l,i} = 0 \right\}.
\end{equation}

Now, we use the homogeneous coordinates to express $\widehat{G}_l$. The nef partition $\widehat{E}_l$ gives a monomial for each $\widehat{G}_l$. They are
\begin{equation}
\prod_{i = 1}^8 \widehat{X}_{l,i}
\end{equation}
Other terms in $\widehat{G}_l$ are determined by the lattice points in $\nabla_l$ as explained above.
$\nabla_l$ has three such points $0, e_{2l-1}$, and $e_{2l}$. The corresponding monomials are:
\begin{equation}
\begin{split}
e_{2l-1} &: \prod_{m = 1}^4 \widehat{X}_{m,2l-1}^2\\
e_{2l} &: \prod_{m = 1}^4 \widehat{X}_{m, 2l}^2
\end{split}
\end{equation}
Then, the most general $\widehat{G}_1, \ldots, \widehat{G}_4$ are (up to isomorphism)
\begin{equation}\label{Ghat}
\begin{split}
\widehat{G}_1 & = \prod_{m = 1}^4 \widehat{X}_{m,1}^2 + \prod_{m = 1}^4 \widehat{X}_{m, 2}^2 - 2 \psi  \prod_{i = 1}^8 \widehat{X}_{1,i}\\
\widehat{G}_2 & = \prod_{m = 1}^4 \widehat{X}_{m,3}^2 + \prod_{m = 1}^4 \widehat{X}_{m, 4}^2 - 2 \psi  \prod_{i = 1}^8 \widehat{X}_{2,i}\\
\widehat{G}_3 & = \prod_{m = 1}^4 \widehat{X}_{m,5}^2 + \prod_{m = 1}^4 \widehat{X}_{m, 6}^2 - 2 \psi  \prod_{i = 1}^8 \widehat{X}_{3,i}\\
\widehat{G}_4 & = \prod_{m = 1}^4 \widehat{X}_{m,7}^2 + \prod_{m = 1}^4 \widehat{X}_{m, 8}^2 - 2 \psi  \prod_{i = 1}^8 \widehat{X}_{4,i}
\end{split}
\end{equation}

$\widehat{V}$ and $\widehat{Y}$ are not smooth. Both have orbifold singularities. To see this, let's consider
$M^\prime$,  the sub-lattice of $M$ that is generated by the vertices $\left\{ \left.\psi_l\right|_{\sigma_i}\right\}$ of $\nabla^*$. The fan $\widehat{\Sigma}$ is also a fan in $M^\prime$ since every cone in $\widehat{\Sigma}$ is a rational cone in $M^\prime$. Therefore, one can construct a toric variety with lattice $M^\prime$ and fan $\widehat{\Sigma}$. Let's call it $V^\prime$. In   \cite{Fulton:1993}, it  is shown
\begin{equation}\label{VprimeQuotient}
V^\prime = \widehat{V} / \left( M / M^\prime \right).
\end{equation}
In our case, $M / M^\prime \simeq (\BZ_2)^3$. It is not a coincidence that $M/M^\prime$ and $G_{torsion}$ have the same form. The actual origin of $G_{torsion}$ is $M/M^\prime$. 
$V^\prime$ has the same homogeneous coordinates $\widehat{X}_{l,i}$ as $\widehat{V}$  and  can be written as a holomorphic quotient:
\begin{equation}
V^\prime \simeq \frac{\BC^{32} - F_{\widehat{\Sigma}}}{G^\prime}
\end{equation}
Group $G^\prime$ is defined similar to $G$. The only difference is that now $n$ in (\ref{GCond}) runs over $N^\prime$, the dual lattice of $M^\prime$.
$G^\prime$ is isomorphic to $\left(\BC^*\right)^{25}$. 

Let $\widehat{Y}^\prime$ be the intersection of $\widehat{G}_1, \ldots, \widehat{G}_4$ in $V^\prime$. From (\ref{VprimeQuotient})
\begin{equation}
\widehat{Y}^\prime = \widehat{Y} / \left( M/M^\prime\right).
\end{equation}
$V^\prime$ and $\widehat{Y}^\prime$ are still singular since $\widehat{\Sigma}$ is not simplicial. However, the singularity of this type can be easily cured by subdividing non-simplicial cones in $\widehat{\Sigma}$. In the gauge linear sigma model point of view, we change Fayet-Iliopoulos parameters to more generic values. Even after subdivision, $V^\prime$ has still orbifold singularities.   Fortunately, $\widehat{Y}^\prime$ misses these singular points. As  we will see later,  the transversality of $\widehat{G}_1, \ldots, \widehat{G}_4$ holds for generic values of $\psi$. Hence, after subdivision, $\widehat{Y}^\prime$ becomes smooth. That implies, for generic points in the moduli space, that all singularities of the mirror are orbifold singularities coming  from the fixed points of $(\BZ_2)^3$.

\subsection{$Z$ out of $\widehat{Y}$}
In this sub-section, we consider the relation between the actual mirror $\widehat{Y}$ and the previously conjectured mirror $Z$. There is a way to  get $\BP^7/(\BZ_4)^3$, the ambient space of $Z$,  out of $\widehat{V}$. Choose 8 edges, $\tau^\prime_i$, from $\widehat{\Sigma}$ as follows:
\begin{equation}
\begin{split}
\tau^\prime_1 = \widehat{\tau}_{4,1} \qquad
\tau^\prime_2 = \widehat{\tau}_{4,2} &\qquad
\tau^\prime_3 = \widehat{\tau}_{1,3} \qquad
\tau^\prime_4 = \widehat{\tau}_{1,4}\\
\tau^\prime_5 = \widehat{\tau}_{2,5}\qquad
\tau^\prime_6 = \widehat{\tau}_{2,6}&\qquad
\tau^\prime_7 = \widehat{\tau}_{3,7}\qquad
\tau^\prime_8 = \widehat{\tau}_{3,8}
\end{split}
\end{equation}
Let $\Sigma^\prime$ be the fan whose edges are $\tau^\prime_i$'s. The toric variety associated to this fan is $\BP^7/(\BZ_4)^3$. It is possible to subdivide non-simplicial cones in $\widehat{\Sigma}$ in such a way  that  every subdivided cone is included in a cone in $\Sigma^\prime$. Then, there is a proper map from the toric variety associated to this subdivision  to $\BP^7/(\BZ_4)^3$ \cite{Fulton:1993}. This map is the blow-up of $\BP^7/(\BZ_4)$ along the singular locus. Furthermore, we get the 4 equations in (\ref{EqsZ}) defining $Z$ from the sets, $\Pi$ and $\Pi^*$.

From this construction, one might wonder if $Z$ corresponds to a point in some corner of the K\"ahler moduli space of $\widehat{Y}$. The answer is ``no". First of all, $\widehat{Y}$ is not a blow-up of $Z$.
In $\widehat{V}$, there are natural liftings of the equation $G_1, \ldots, G_4$ in (\ref{EqsZ}). It turns out that they are not $\widehat{G}_1, \ldots, \widehat{G}_4$, but $\widehat{G}_l$ times some monomials.
Each monomial that multiplies $\widehat{G}_l$ involves only the homogeneous coordinates that corresponds to the exceptional divisors. These divisors are represented by edges of $\widehat{\Sigma}$ that are not edges of $\Sigma^\prime$. The 4 equations obtained this way define a singular variety. This  variety  is reducible and contains $\widehat{Y}$ as an irreducible component. Only when we shrink the exceptional divisors to size zero, it becomes irreducible. However, it still contains singularities. This is  the origin of $Z$'s singularities where the transversality fails.

The image of $\widehat{Y}$ under the proper map is $Z$. Hence,  it is interesting to see what happens to $\widehat{Y}$ when we "blow down" the exceptional divisors. The gauge linear sigma model describing $Z$ has non-compact vacua. Recall this model has fields $P_l$ that multiply $G_l$ to comprise the superpotential. The theory has the flat direction along $P_l$'s at points on $Z$ where the transversality of $G_l$ fails. However, this is not what happens in the gauge linear sigma model of $\widehat{Y}$. The flat directions will be replaced by compact manifolds parameterized by $P_l$'s and the fields corresponding to the exceptional divisors. 

To illustrate this point, let's consider a toy model. Take $\BP^4$ and blow up a point $[0,0,0,0,1]$.
The gauge linear sigma model describing this blown up space has 6 fields, $X_1, \ldots, X_5, T$ and two $U(1)$'s. The charges of the fields are:
\begin{equation}
\begin{array}{c|*{6}c}
& X_1 & X_2 & X_3 & X_4 & X_5  & T\\
\hline
U(1)_1 & 1 & 1 & 1& 1& & -1\\
U(1)_2 &  & & & & 1 & 1 
\end{array}
\end{equation}
To have a Calabi-Yau hypersurface, the defining equation $G$ must have charge $(3,2)$ and is in the following form:
\begin{equation}
G(X_1, \ldots, X_5, T) = G_3(X_1, \ldots, X_4) X_5^2 + G_4(X_1, \ldots, X_4) X_5 T + G_5(X_1, \ldots, X_4) T^2
\end{equation}
where $G_3, G_4$, and $G_5$ are homogeneous polynomials of $X_1, \ldots, X_4$ with degree 3, 4, and 5, respectively. For generic enough $G_3,  G_4$ and $G_5$, the hypersurface is a smooth Calabi-Yau manifold. At the point $X_1 = \ldots = X_4 = 0$, $G = dG = 0$. However, this point is excluded by $D$-term conditions:
\begin{equation}\label{Dterms}
\begin{split}
r_1 & = \abs{X_1}^2 + \ldots + \abs{X_4}^2 -\abs{T}^2 - 3\abs{P}^2 \\
r_2 & = \abs{X_5}^2 + \abs{T}^2 - 2\abs{P}^2
\end{split}
\end{equation}
where $P$ is a field introduced for $G$ and, $r_1$ and $r_2$ are Fayet-Iliopoulos parameters. To represent the hypersurface in the blown up space, $r_1$ and $r_2$ must be taken positive.
Now, let's blow down and see what happens. Blowing down corresponds to changing the value of $r_1$ to a negative number while keeping $r_1 + r_2$ positive. If there were no $P$  in the first equation of (\ref{Dterms}), we would obtain $\BP^4$. The first equation would determine the non-zero absolute value of $T$ and the phase of $T$ would be gauged away by  $U(1)_1$. Since the point $X_1 = \ldots = X_4 = 0$ is  allowed now, $G$ becomes singular. That would allow $P$ to take any value making the space of vacua non-compact.  This is very similar to our situation. However, the presence of $P$ in the first equation of (\ref{Dterms}) makes the difference. Even though $G$ is singular, the space of vacua remains compact since the absolute value of $P$ is bounded by the first equation in (\ref{Dterms}). Now, at the  point $X_1 = \ldots X_4 = 0$, there is $\BW^1_{(1,3)}$ parameterized by $T$ and $P$. 
So far, our discussion has been classical. Of course, the quantum correction will change the shape of the vacua. However, we expect the quantum correction to be not so dramatic  that the space of vacua remains compact.

\section{Monomial-divisor mirror map}
In section \ref{Beauville}, we saw $Q$ acts on the space of quadrics in $\BP^7$. Hence, there is an induced $Q$-action on the complex structure moduli space, $\mathcal{C}_Y$, of $Y$. To define the Beauville manifold, we demand the complex structure to be in subset $\mathcal{C}_Y^0$ of $\mathcal{C}_Y$ that corresponds to choices of 4 quadrics, one from each irreducible representation of $Q$. With this choice of the complex structure, $Q$ acts freely on $Y$ as a subgroup of the automorphisms.  
On the mirror side, we expect there is the mirror $Q$-action on the K\"ahler moduli space $\mathcal{K}_{\widehat{Y}}$ of $\widehat{Y}$. To define the mirror of the Beauville manifold, we need to tune $\widehat{Y}$'s K\"ahler parameters so that the K\"ahler class is fixed by $Q$. Then, we hope to find a compatible $Q$-action on $\widehat{Y}$ itself and take the quotient of $\widehat{Y}$ by the $Q$-action. 

To find the $Q$-action  on $\mathcal{K}_{\widehat{Y}}$, we, presumably,  need to solve the mirror map $\mu: \mathcal{C}_Y \to\mathcal{K}_{\widehat{Y}}$. However, with 65 parameters, it is practically impossible. (Here, we are talking about the mirror map between the complex structure of $Y$ and the K\"ahler class of $\widehat{Y}$.)
Instead of solving the mirror map directly, we will use the monomial-divisor mirror map  \cite{Aspinwall:1993rj}. It is the differential of the mirror map and is valid only in large radius limit.
However, without knowledge of the actual mirror map, we do not know which part of the complex structure moduli space is mapped to the large radius limit. We will have to conjecture the large radius limit is the region of $\mathcal{C}_Y$ where the monomial-divisor map is well defined.   It turns out that in this conjectured large radius limit, there are $Q$ fixed points. 
Hence, we tune the K\"ahler parameters to one of those points, and apply the monomial-divisor map to find out $Q$-action on the mirror. In this section, we describe the tangent spaces of $\mathcal{K}_{\widehat{Y}}$ and $\mathcal{C}_Y$ and the monomial-divisor mirror map following   \cite{Aspinwall:1993rj}. The authors of   \cite{Aspinwall:1993rj} only considered Calabi-Yau manifolds that are hypersurfaces in toric varieties. We will need to extend the argument a little bit to suit complete intersection Calabi-Yau cases.

\subsection{Divisors}
We describe the tangent space of $\mathcal{K}_{\widehat{Y}}$.  
First, we resolve all singularities of $\widehat{V}$.\footnote{It is not necessary to resolve all singularities since some miss $\widehat{Y}$. We only need to resolve singularities of $\widehat{Y}$. However, it will turn out to be more useful in understanding the monomial-divisor map to resolve all.}  We add new edges to the fan $\widehat{\Sigma}$ and subdivide cones so that the resulting fan is simplicial. In this way, we get the blown up $\widetilde{V}$ of $\widehat{V}$. There is an induced blown up $\widetilde{Y}$ of $\widehat{Y}$, too.  The new edges will be generated by lattice points of $\nabla^*$. Since $\nabla^*$ is reflexive, it contains only one interior lattice point,  $0$: the other lattice points are on faces of $\nabla^*$. Let's call $\Xi$ the set of these points:
\begin{equation}
\Xi =  \nabla^* \cap M - \{0\}.
\end{equation}
It turns out that every lattice point in $\nabla^*$ is in one of $\Delta_l$'s. This is a necessary condition for the monomial-divisor mirror map to be well defined for complete intersection Calabi-Yau cases. We do not know if this condition is generally met for other cases. $\Xi$ contains 140 points: $\omega_i + \omega_j - \omega_{2l-1}-\omega_{2l}$. Label them $\xi_{l,i,j}$ and denote by $\widetilde{D}_{l,i,j}$ the  associated divisors. There is an isomorphism between $\BZ^\Xi$, the set of integer-valued functions on $\Xi$, and $\mathrm{WDiv}_T(\widetilde{V})$, the set of $T$-stable Weil divisors in $\widetilde{V}$. Given  an integer-valued function $\phi$ on $\Xi$, one can construct a Weil divisor, $\sum \phi(\xi_{l,i,j})\widetilde{D}_{l,i,j}$.  The Chow group $A_6(\widetilde{V})$ is the quotient of $\mathrm{WDiv}_T(\widetilde{V})$ by linear equivalence. As $M$ represents the set of meromorphic functions on $\BP^7$, meromorphic functions on $\widetilde{V}$ will be represented by elements in $N$.
For $n \in N$, we denote the corresponding function by $\widetilde{\chi}^n$.
Two linearly equivalent Weil divisors, $W, W^\prime$ are related to each other by
\begin{equation}
W - W^\prime = \sum \langle n, \xi_{l,i,j} \rangle \widetilde{D}_{l,i,j}
\end{equation}
for some $n \in N$. 
Equivalently, consider embedding $\mathrm{ad}_\Xi : N  \to \BZ^\Xi$ by sending $n \in N$ to the function $\mathrm{ad}_\Xi(n)$ defined by $\mathrm{ad}_\Xi(n) : \xi_{l,i,j} \mapsto \langle n, \xi_{l,i,j} \rangle$. Then, 
\begin{equation}
\begin{split}
A_6(\widetilde{V}) & \simeq \BZ^\Xi / N \\
\coho{1,1}{\widetilde{V}} &\simeq \left(\BZ^\Xi / N \right) \otimes \BC.
\end{split}
\end{equation}

The tangent space of $\mathcal{\widehat{K}}$ is not $\coho{1,1}{\widetilde{V}}$ but $\coho{1,1}{\widetilde{Y}}$. Let's consider $\mathrm{WDiv}_T(\widetilde{Y})$. Generally, a divisor in $\widetilde{V}$ induces a divisor in $\widetilde{Y}$. However, some  divisors in $\mathrm{WDiv}_T(\widetilde{V})$ do not intersect $\widetilde{Y}$. We need to exclude those divisors. 
Detailed investigation of $\widehat{\Sigma}$ shows that the following 68 divisors do not intersect $\widetilde{Y}$ for $l, m = 1, \ldots, 4$, $i = 1, \ldots, 8$ and $i \neq 2l-1, 2l$ and $m \neq l$:
\begin{equation}
\widetilde{D}_{l,2l-1,i}\quad \widetilde{D}_{l,2l,i}\quad \widetilde{D}_{l,2m-1, 2m}.
\end{equation}
Let $\Xi^0$ be the set of lattice points representing divisors that do intersect $\widetilde{Y}$. $\Xi^0$ contains 72 lattice points. As before, $\mathrm{WDiv}_T(\widetilde{Y}) \simeq \BZ^{\Xi^0}$, and the linear equivalence is given by the embedding of $N$ in $\BZ^{\Xi^0}$. Therefore,\footnote{Sometimes, the toric divisors do not generate the entire Chow group. In our case, we know that $h^{1,1}(\widetilde{Y}) = 65$ by mirror symmetry. Since $\dim \left[\left(\BZ^{\Xi^0} / N\right) \otimes \BC\right] = 65$, we conclude that the above expression is indeed correct}
\begin{equation}\label{TK}
\coho{1,1}{\widetilde{Y}} \simeq \left(\BZ^{\Xi^0} / N \right) \otimes \BC.
\end{equation}

\subsection{Monomials}
The tangent space of $\mathcal{C}_Y$ is $\coho{2,1}{Y}$. The simplest way to deform the complex structure of $Y$ is to perturb the defining equations, $G_1, \ldots, G_4$. Generally speaking, there might be deformations that cannot be realized in this way \cite{Green:1987rw}. In our case, the dimensional analysis tells us there are no such deformations.
As we explained in the previous section, the monomials in equation $G_l$ are represented by the lattice points in $\Delta_l$. Let $\mathcal{A}$ be the space of 4 quadrics in $\BP^7$. Then,
\begin{equation}\label{AToric}
\mathcal{A} = \bigotimes_{l = 1}^4\BC^{\Delta_l \cap M}
\end{equation} 
Here, each factor $\BC^{\Delta_l \cap M}$ represents the coefficients of the monomials in $G_l$.
Not every point of $\mathcal{A}$ will give distinct choice of the complex structure. 
Linear transformations of $G_l$'s will define the same manifold $Y$. Also, linear transformations of $X_i$'s will give manifolds isomorphic to the original manifold $Y$.   
From the gauge sigma model analysis, we know that infinitesimal changes in the superpotential $W$ that are generated by its derivatives  will give an isomorphic theory.
Since $W = P_1 G_1 + \ldots + P_4 G_4$,  such changes are
\begin{equation}
\delta W = \alpha_{m,l} P_l \frac{\partial W}{\partial P_m} + \beta_{j,i} X_i \frac{\partial W}{\partial X_j}.
\end{equation}
They are generated by the infinitesimal linear transformations of $P_l$ and $X_i$.
Some elements of $\mathrm{GL}(4,\BC) \otimes \mathrm{GL}(8, \BC)$ act trivially on $\mathcal{A}$. They are 
\begin{equation}\label{CommonFactor}
\begin{split}
G_l &\mapsto \lambda^{-2} G_l\\
X_i & \mapsto \lambda X_i
\end{split}
\end{equation}
for some $\lambda \in \BC^*$.\footnote{As we will see later, at special points in $\mathcal{A}$, $\mathrm{GL}(4,\BC)\otimes\mathrm{GL}(8, \BC)$ may have additional elements that act trivially. But, there are only a finite number of them and they do not change our arguments below.}  Hence, the actual group acting on $\mathcal{A}$ is
\begin{equation}
\mathcal{G} = \left( \mathrm{GL}(4, \BC) \otimes \mathrm{GL}(8, \BC)\right) / \BC^*.
\end{equation}
Therefore, we find that
\begin{equation}
\mathcal{C}_Y \simeq \mathcal{A} /\left( \mathcal{G} \otimes \Gamma\right)
\end{equation}
where $\Gamma$ is the discrete group that consists of diffeomorphisms of $Y$  not connected to the identity, hence not captured in the above discussion.

 We would like to describe the tangent space of $\mathcal{C}_Y$ using toric geometry.
Unfortunately, not every element of $\mathcal{G}$ is compatible with the toric description. 
First, some elements of $\mathrm{GL}(8, \BC)$ are not compatible with the $T$-action, the torus action on the toric variety. As we have seen in the previous section, $T$-equivariant Cartier divisors are monomials. Generally, linear transformations  of $X_i$'s map monomials to polynomials. Such linear transformations are not compatible  with the $T$-action.
Second, $G_l$'s are represented by polyhedrons $\Delta_i$ and it is not clear how to represent linear transformations of $G_l$'s in terms of polyhedrons. 
However, there are two kinds elements of $\mathcal{G}$ that are compatible with $T$. They are scalings
\begin{equation}
\begin{split}
G_l &\mapsto \rho_l G_l \quad\text{(no sum)}\\
X_i &\mapsto \lambda_i X_i \quad\text{(no sum)}
\end{split}
\end{equation}
and permutations:
\begin{equation}
\begin{split}
G_l &\mapsto G_{\pi(l)}\\
X_i &\mapsto X_{\pi^\prime(i)}
\end{split}
\end{equation}
where $\pi, \pi^\prime$ are elements of $S_4$ and $S_8$, the permutation group of 4 and 8 objects, respectively.
In finding the toric description of the tangent space of $\mathcal{C}_Y$, only scalings will play an important role. Taking account of the elements acting trivially on $\mathcal{A}$, we find that scalings comprise a subgroup $\mathcal{G}_T$ isomorphic to 
\begin{equation}\label{GT}
T \otimes (\BC^*)^4. 
\end{equation}
Since only part of $\mathcal{G}$ is realized in the  toric description, we will use the rest of $\mathcal{G}$ to fix some coefficients of the monomials in $G_l$. This is a tricky procedure. Generally, the choice of monomials whose  coefficients will be fixed depends on where you are on the moduli space. Since we are interested in finding the monomial-divisor mirror map, we would like to be in the large radius limit. However, without knowledge of the actual mirror map, we do not know which part of $\mathcal{A}$ is mapped to the large radius limit. We need to guess. Mirror symmetry tells us that the tangent space of $\mathcal{C}_Y$ in the large radius limit is isomorphic to 
\begin{equation}
\left(\BZ^{\Xi^0} / N \right) \otimes \BC.
\end{equation}
Notice the dual role the lattice points in $\Delta_l$ play here. The lattice point $\xi_{l,i,j} = \omega_i + \omega_j - \omega_{2l-l} - \omega_{2l}$ represents the monomial $X_i X_j$ in $G_l$. It also represents toric divisor $\widetilde{D}_{l,i,j}$ of $\widetilde{V}$ on the mirror side. There is a  subtlety. While $0$ in $\Delta_l$ represents monomial $X_{2l-1} X_{2l}$ of $G_l$, there is no divisor corresponding to this point. Also, unlike other points, $0$ represents 4 monomials, one from each $G_l$ since it is the only common point of $\Delta_1, \ldots, \Delta_4$. Using the fact that $\Xi = \bigcup_{l = 1}^4 \left(\Delta_l \cap M\right) - \{0\}$, we have
\begin{equation}
\mathcal{A} \simeq  \BC^\Xi \otimes \BC^4
\end{equation}
where the extra factor $\BC^4$ represents the coefficient of the 4 monomials that the point $0$ represents.
This $\BC^4$ will be cancelled by $(\BC^*)^4$ in (\ref{GT}). Remember that the factor $(\BC^*)^4$ represents scaling:
\begin{equation}
G_l \mapsto \rho_l G_l \quad \text{(no sum)}.
\end{equation} 
We can use this scaling to set the coefficient of $X_{2l-1} X_{2l}$ in $G_l$ to 1 as long as the original coefficients are non-zero. There are other coefficients to fix. Some of divisors in $\widetilde{V}$ do not intersect $\widetilde{Y}$. 
Only divisors corresponding to the points in $\Xi^0$ intersect $\widetilde{Y}$. Therefore, it is tempting to fix the coefficients of the monomials corresponding to the points in $\Xi$ which are not in $ \Xi^0$. There are 68 such points. Since the dimension of $\mathcal{G} / \mathcal{G}_T$ is 68, it is plausible that we could fix these 68 coefficients. This time, we want to set these coefficients to zero so that $\mathcal{G}_T$ does not change them.
Hence, we conjecture that  in the large radius limit of $\mathcal{A}$, there is an element of  $\mathcal{G}$ that we can use to set 
\begin{itemize}
\item the coefficient of $X_{2l-1} X_{2l}$ of $G_l$ to 1
\item the coefficient of  monomials:
\begin{equation}
X_{2l-1} X_i,\quad X_{2l}X_i, \quad X_{2m-1}X_{2m} \qquad \text{$i \neq 2l-1, 2l$ and $m \neq l$}
\end{equation}
of $G_l$ to zero .
\end{itemize}

Once we fixed the coefficients, the conjectured large radius limit of $\mathcal{C}_Y$ looks like
\begin{equation}
\BC^{\Xi^0} / T
\end{equation}
and the tangent space $\coho{2,1}{Y}$ is 
\begin{equation}\label{TC}
\BC^{\Xi^0} / \left(N\otimes \BC\right).
\end{equation}
Here, we have used the fact that the tangent space of $T$ is $N\otimes \BC$.\footnote{This can be a little confusing since $T = N \otimes_\BZ \BC^*$.  The group multiplication is the addition in $N$ and the multiplication in $\BC^*$. Now, we can represent $\BC^*$ by $\exp \BC$. Then, $T =  \exp N \otimes \BC$.}
Now, the monomial-divisor mirror map is evident since 
\begin{equation}
\left(\BZ^{\Xi^0}/ N \right) \otimes \BC \simeq \BC^{\Xi^0} / \left(N\otimes \BC\right)
\end{equation}

\section{Mirror of the Beauville manifold}
We will construct  the mirror of the Beauville manifold by taking a quotient $\widetilde{X} = \widetilde{Y} / Q$. To do so, first, we need to find how $Q$ acts on $\widetilde{Y}$.

\subsection{Mirror $Q$-action} 
We tune our parameters to one of the $Q$-fixed points of $\mathcal{C}_Y$. 
It amounts to choosing 4 quadrics, one from each one-dimensional irreducible representation of $Q$. 
$Y$ defined with this choice of the quadrics, $G_1, \ldots, G_4$, is invariant under $Q$ because the induced $Q$-action on quadrics will transform $G_l$'s to some linear combinations of them.  We can embed $Q$ into $\mathrm{GL}(4, \BC)$ such that this embedded $Q$ takes the transformed $G_l$'s back to the original. 
In other words, at these points, there are additional elements of  $\mathrm{GL}(4, \BC)\otimes\mathrm{GL}(8,\BC)$ that act trivially on $\mathcal{A}$ and they comprise a subgroup isomorphic to $Q$. Actual embedding of $Q$ is determined by the choice of the homogeneous coordinates and the quadrics that give the same $Y$. We want this embedding to be compatible with the toric description. Since elements of $\mathcal{G}$ compatible with the toric description are  scalings and permutations, we need to find the choice of the homogeneous coordinates and the quadrics  that embeds $Q$ into the scalings and permutations. 

One might wonder why it matters because every choice of the homogeneous coordinates and the quadrics is related to each other by linear transformations and any choice is as good as any other. However, this is not true. First, linear transformations of the homogeneous coordinates and the quadrics will map scalings and permutations to other linear transformations that are not compatible with the toric description. Second, to get the toric description of the tangent space of $\mathcal{C}_Y$, we have fixed some coefficients of the quadrics. If we linear-transform the homogeneous coordinates or the quadrics, it will change the fixed coefficients and ruin the toric description and hence, also the monomial-divisor mirror map.

It turns out that the right choice will embed $Q$ into permutations. We choose the homogeneous coordinates such that $Q$-action is given by:
\begin{equation}
g_1 \cdot X_{g_2} = X_{g_1 g_2}
\end{equation}
where $g_1, g_2 \in Q$ and we have relabeled the homogeneous coordinates $X_g$. To explain the choice of the quadrics, we note that $V_4 = V_1 \oplus V_I \oplus V_J \oplus V_K$ is the regular representation of  the abelianization $Q / [Q, Q] = \BZ_2 \times \BZ_2$. Elements of $Q / [Q, Q]$ are the cosets of the normal subgroup, $[Q, Q] =   \{1, -1\}$. We denote by $[g]$ the coset $g \cdot [Q, Q]$ for each $g \in Q$. Then $[g] = [-g]$ and the multiplication law is given by $[g_1][g_2] = [g_1 g_2]$.
We label the quadrics, $G_{[g]}$ and choose them such that $Q$ acts on them as 
\begin{equation}
g_1 \cdot G_{[g_2]} = G_{[g_1 g_2]}
\end{equation}
where $g_1, g_2 \in Q$.
With this choice of the homogeneous coordinates and the quadrics, it is clear that $Q$ acts on them as permutations.
Now, let's write down the most general $G_{[g]}$'s and see if there is a fixed point in the conjectured large radius limit. The most general $G_{[g]}$'s are 
\begin{equation*}
\begin{split}
G_{[1]} & = t_1 X_1X_{-1} + t_2 X_I X_{-I} + t_3 X_J X_{-J} + t_4 X_K X_{-K}\\
&\qquad + t_5(X_1^2 + X_{-1}^2) + t_6(X_I^2 + X_{-I}^2) + t_7 (X_J^2 + X_{-J}^2) + t_8 (X_K^2 + X_{-K}^2)\\
&\qquad + t_9 (X_1X_I + X_{-1} X_{-I}) + t_{10}(X_1X_J + X_{-1}X_{-J})+ t_{11}(X_1 X_K + X_{-1} X_{-K})\\
&\qquad + t_{12} (X_1X_{-I} + X_{-1}X_{I}) +t_{13}(X_1 X_{-J} + X_{-1} X_J) + t_{14}(X_1 X_{-K} + X_{-1}X_K)\\
&\qquad + t_{15}(X_I X_J + X_{-I}X_{-J}) + t_{16}(X_J X_K + X_{-J}X_{-K})+ t_{17}(X_K X_I + X_{-K}X_{-I})\\
&\qquad   + t_{18}(X_I X_{-J} + X_{-I}X_J) + t_{19}(X_J X_{-K}+ X_{-J} X_K)   + t_{20}(X_K X_{-I}+ X_{-K} X_I) \\
G_{[I]} & = t_1 X_IX_{-I} + t_2 X_1 X_{-1} + t_3 X_K X_{-K} + t_4 X_J X_{-J}\\
&\qquad + t_5(X_I^2 + X_{-I}^2) + t_6(X_1^2 + X_{-1}^2) + t_7 (X_K^2 + X_{-K}^2) + t_8 (X_J^2 + X_{-J}^2)\\
&\qquad + t_9 (X_1X_{-I} + X_{-1} X_I) + t_{10}(X_IX_K + X_{-I}X_{-K})+ t_{11}(X_I X_{-J} + X_{-I} X_J)\\
&\qquad + t_{12} (X_1X_I + X_{-1}X_{-I}) +t_{13}(X_K X_{-I} + X_{-K} X_I) + t_{14}(X_I X_J + X_{-I}X_{-J})\\
&\qquad + t_{15}(X_1 X_{-K} + X_{-1}X_K) + t_{16}(X_J X_{-K} + X_{-J}X_K)+ t_{17}(X_1 X_J + X_{-1}X_{-J})\\
&\qquad   + t_{18}(X_1 X_K + X_{-1}X_{-K}) + t_{19}(X_J X_K+ X_{-J} X_{-K})   + t_{20}(X_1 X_{-J}+ X_{-1} X_J)
\end{split}
\end{equation*}
\begin{equation}\label{MostGeneralG}
\begin{split}
G_{[J]} & = t_1 X_JX_{-J} + t_2 X_K X_{-K} + t_3 X_1 X_{-1} + t_4 X_I X_{-I}\\
&\qquad + t_5(X_J^2 + X_{-J}^2) + t_6(X_K^2 + X_{-K}^2) + t_7 (X_1^2 + X_{-1}^2) + t_8 (X_I^2 + X_{-I}^2)\\
&\qquad + t_9 (X_JX_{-K} + X_{-J} X_K) + t_{10}(X_1X_{-J} + X_{-1}X_J)+ t_{11}(X_I X_J + X_{-I} X_{-J})\\
&\qquad + t_{12} (X_JX_K + X_{-J}X_{-K}) +t_{13}(X_1 X_J + X_{-1} X_{-J}) + t_{14}(X_I X_{-J} + X_{-I}X_J)\\
&\qquad + t_{15}(X_1 X_K + X_{-1}X_{-K}) + t_{16}(X_1 X_{-I} + X_{-1}X_I)+ t_{17}(X_K X_{-I} + X_{-K}X_I)\\
&\qquad   + t_{18}(X_1 X_{-K} + X_{-1}X_K) + t_{19}(X_1 X_I+ X_{-1} X_{-I})   + t_{20}(X_K X_I+ X_{-K} X_{-I}) \\
G_{[K]} & = t_1 X_KX_{-K} + t_2 X_J X_{-J} + t_3 X_I X_{-I} + t_4 X_1 X_{-1}\\
&\qquad + t_5(X_K^2 + X_{-K}^2) + t_6(X_J^2 + X_{-J}^2) + t_7 (X_I^2 + X_{-I}^2) + t_8 (X_1^2 + X_{-1}^2)\\
&\qquad + t_9 (X_JX_K + X_{-J} X_{-K}) + t_{10}(X_KX_{-I} + X_{-K}X_I)+ t_{11}(X_1 X_{-K} + X_{-1} X_K)\\
&\qquad + t_{12} (X_JX_{-K} + X_{-J}X_K) +t_{13}(X_K X_I + X_{-K} X_{-I}) + t_{14}(X_1 X_K + X_{-1}X_{-K})\\
&\qquad + t_{15}(X_I X_{-J} + X_{-I}X_J) + t_{16}(X_1 X_I + X_{-1}X_{-I})+ t_{17}(X_1 X_{-J} + X_{-1}X_J)\\
&\qquad   + t_{18}(X_I X_J + X_{-I}X_{-J}) + t_{19}(X_1 X_{-I}+ X_{-1} X_I)   + t_{20}(X_1X_J+ X_{-1} X_{-J})
\end{split}
\end{equation}
We choose the following nef partition:
\begin{equation}
\Delta_{[g]} = \{ e_g, e_{-g} \}
\end{equation}
where $e_g$ is the vertex of $\Delta^*$ that represents the divisor $D_g = \{ X_g = 0 \}$.
Then, there are $Q$-fixed points in the large radius limit.  They are points with $t_1 = 1, t_2 = t_3 = t_4 = t_5 = t_9 = t_{10} = t_{11}= t_{12} = t_{13} = t_{14} = 0$.

Now, we tune our parameters to one of these points and apply the monomial-divisor mirror map to find the mirror $Q$-action on $\widetilde{Y}$.
Actually, it turns out that $\widetilde{V}$ has a $Q$-action too and it is easy to identify. To do so, we introduce the homogeneous coordinates on $\widetilde{V}$
\begin{equation}
\widetilde{X}_{[g_1],g_2,g_3}
\end{equation}
where $[g_1] \in Q/[Q,Q]$ and $g_2, g_3 \in Q$. Since we have relabeled the homogeneous coordinates and the quadrics, we relabel everything on the mirror side accordingly. 
As usual, $\widetilde{X}_{[g_1],g_2,g_3}$ is determined up to a multiplicative constant by $\widetilde{D}_{[g_1],g_2,g_3} = \left\{ \widetilde{X}_{[g_1],g_2,g_3} = 0 \right\}$. $\widetilde{A}$ also has the holomorphic quotient description
\begin{equation}
\widetilde{V}  \simeq \frac{\BC^{140} - F_{\widetilde{\Sigma}}}{\widetilde{G}}.
\end{equation}
The group $\widetilde{G}$ is defined similar to $G$:
\begin{equation}\label{GaugeXtilde}
\widetilde{X}_{[g_1], g_2, g_3} \mapsto \Lambda_{[g_1], g_2, g_3} \widetilde{X}_{[g_1], g_2, g_3} \quad\text{(no sum)}.
\end{equation}
Here, $\Lambda_{[g_1], g_2, g_3}$'s are non-zero complex numbers with 
\begin{equation}
1 = \prod_{[g_1] \in Q/[Q,Q] \atop g_2, g_3 \in Q} \Lambda_{[g_1], g_2, g_3}^{\langle n, \xi_{[g_1], g_2, g_3} \rangle} \qquad \forall n \in N.
\end{equation}
It is easy to show $\widetilde{G} \simeq (\BC^*)^{133}$.

Since we are at one of the $Q$-fixed points  of $\mathcal{A}$, we have $Q$-action on its tangent space $T\mathcal{A}$. Note 
\begin{equation}
T\mathcal{A} = \BC^\Xi \oplus \BC^4.
\end{equation}
Here, the term $\BC^4$ corresponds to the coefficients of the 4 monomials that the point $0$ represents and is invariant under the $Q$-action. $Q$ permutes these 4 monomials. 
Each point $\xi \in \Xi$ represents both a monomial in one of the quadrics and a $T$-stable Weil divisor of $\widetilde{V}$. Since we know how $Q$ permutes monomials, we can deduce the $Q$-action on $\mathrm{WDiv}_T(\widetilde{V})$:
\begin{equation}
g \cdot \widetilde{D}_{[g_1], g_2, g_3} = \widetilde{D}_{[g g_1], g g_2, g g_3}.
\end{equation}
In terms of the homogeneous coordinates
\begin{equation}
g \cdot \widetilde{X}_{[g_1], g_2, g_3} = \widetilde{X}_{[g g_1], g g_2, g g_3}.
\end{equation}
Since this $Q$-action maps polyhedron $\nabla^*$ to itself, it is consistent with the construction of the toric variety $\widetilde{V}$. 

\subsection{Mirror of the Beauville manifold}
To define $\widetilde{X} = \widetilde{Y} / Q$, we need to show $\widetilde{Y}$ is invariant under $Q$ and $Q$ acts freely on $\widetilde{Y}$.
The easiest way of showing this is to go back to $\widehat{Y}$ and work there. $Q$ permutes the vertices of $\nabla^*$. Therefore, $\widehat{V}$ has the $Q$-action too:
\begin{equation}
g \cdot \widehat{X}_{[g_1], g_2} = \widehat{X}_{[g g_1], g g_2}
\end{equation}
One can easily check that this $Q$-action permutes $\widehat{G}_{[g]}$'s:
\begin{equation}
\widehat{G}_{[g_1]}( g \cdot \widehat{X}) = \widehat{G}_{[g g_1]}( \widehat{X}). 
\end{equation}
This implies $\widehat{Y}$ is invariant under $Q$. Now, we would like to show that $Q$ acts freely on $\widehat{Y}$. Suppose there is a point $p \in \widehat{Y}$ that is fixed by some non-identity elements of $Q$.  Then, $p$ is also fixed by $-1 \in Q$. Note that the values of the homogeneous coordinates are fixed by -1 up to a gauge transformation.  Hence, at $p$, we have
\begin{equation}\label{FixedCond}
\widehat{X}_{[g_1], g_2} = \Lambda_{[g_1], g_2}\widehat{X}_{[g_1], -g_2} \quad\text{(no sum)}
\end{equation}
for some $\Lambda_{[g_1], g_2} \in \BC^*$ satisfying (\ref{GCond}).
Plugging the above equation in (\ref{GCond}), we get 
\begin{equation}
\prod_{[g^\prime] \in Q/[Q,Q]} \widehat{X}_{[g^\prime],g}^2 = \prod_{g^\prime \in Q/[Q,Q]}\widehat{X}_{[g^\prime],-g}^2
\end{equation}
for all $g \in Q$. 
With (\ref{Ghat}), this implies
\begin{equation}\label{EqFixedPts}
(\psi^8 - 1) \prod_{[g_1] \in Q/[Q,Q] \atop g_2 \in Q} \widehat{X}_{[g_1] g_2}^2 = 0.
\end{equation}
By closely investigating $\widehat{\Sigma}$, one can show\footnote{One way to show this is the following. In   \cite{Borisov:1993}, it is shown that $\nabla = \nabla_1 + \ldots + \nabla_4$. With this, one can easily identify the full dimensional cones in $\widehat{ \Sigma}$. The above follows from the the fact that divisors have common points only if they all belong to a same full dimensional cone in $\widehat{\Sigma}$ \cite{Fulton:1993}.}
\begin{itemize}
\item $X_{[g], g}$ and $X_{[g], -g}$ do not intersect $\widehat{G}_{[g]}$ (and hence $\widehat{Y}$).
\item For $[g_1] \neq [g_2]$,  $\left\{\widehat{X}_{[g_1], g_2} =  \widehat{X}_{[g_1], -g_2} = 0\right\}$ does not intersect $\widehat{G}_{[g_2]}$ (and hence $\widehat{Y}$).
\end{itemize}
Together with (\ref{FixedCond}), it is easy to see no $\widehat{X}_{[g_1], g_2}$ vanishes at $p$.
Therefore, there is no solution to (\ref{EqFixedPts}) on $\widehat{Y}$ unless $\psi^8 = 1$. As we will see later,  $\psi^8 = 1$ is the conifold point. For generic value of $\psi$, there is no such $p$, and  $Q$ acts freely on $\widehat{Y}$. Since $\widetilde{Y}$ is the blow-up of $\widehat{Y}$, we conclude that $\widetilde{Y}$ is invariant under $Q$ and the $Q$-action on it is free too. 
We take the quotient of $\widetilde{Y}$ by $Q$ to define the mirror $\widetilde{X}$ of the Beauville manifold. 

Let's take a close look at the moduli spaces of $X$ and $\widetilde{X}$. As in the case of $Y$, the complex structure moduli space, $\mathcal{C}_X$, of $X$ can be described as a quotient of the space, $\mathcal{A}_X$ of 4 quadrics defining $X$ by the group $\mathcal{G}_X$ generating the manifolds that are isomorphic to $X$. The most general 4 quadrics defining $X$ have been written in (\ref{MostGeneralG}). From that, 
\begin{equation}
\mathcal{A}_X \simeq \BC^{20}
\end{equation}
Now, we claim that
\begin{equation}
\mathcal{G}_X  \simeq (\BC^*)^7 \otimes \mathrm{GL}(2,\BC) \otimes \Gamma_X
\end{equation}
 where $\Gamma_X$ is a discrete group whose elements are diffeomorphisms of $X$ not connected to the identity.
To see this, notice \footnote{In the previous subsection, we could have used the homogeneous coordinates and the quadrics that make this decomposition manifest. This was the choice Beauville originally used in \cite{Beauville:1995}. However, this choice would have given no fixed points in the large radius limit.}
\begin{equation}
\begin{split}
V_8 &= V_1 \oplus V_I \oplus V_J \oplus  V_K \oplus 2 V_2\\
V_4 & = V_1 \oplus V_I \oplus V_J \oplus V_K.
\end{split}
\end{equation} 
From this decomposition, it is clear that the factor $(\BC^*)^4 \otimes \mathrm{GL}(2, \BC)$ comes from linear transformations of $X_g$'s and the factor $(\BC^*)^4$ from linear transformations of $G_{[g]}$'s. As before, they have the subgroup $\BC^*$ that acts trivially on $\mathcal{A}_X$ proving the claim.  
Out of $\mathcal{G}_X$, only $\BC^*$ has the toric description. It corresponds to the overall scaling of the homogeneous coordinates. With this $\BC^*$, we set $t_1$ in (\ref{MostGeneralG}) to $1$. The rest of $\mathcal{G}_X$ will be used to set $t_2,  t_3,  t_4,  t_5,  t_9,  t_{10},  t_{11},  t_{12},  t_{13},  t_{14}$ to $0$ (in the large radius limit). Therefore, the tangent space of $\mathcal{C}_X$ in the large radius limit is
\begin{equation}
T\mathcal{C}_X = \BC^9
\end{equation}
where $\BC^9$ is parameterized  by 9 $t_i$'s that are not fixed.

The tangent space of K\"ahler moduli space $\mathcal{K}_{\widetilde{X}}$ of the mirror $\widetilde{X}$ is 
\begin{equation}
T\mathcal{K}_{\widetilde{X}} = \left(\mathrm{WDiv}_T(\widetilde{X}) / \text{linear equivalence} \right)\otimes \BC.
\end{equation}
For a given $T$-stable Weil divisor of $\widetilde{X}$, one finds a $Q$-invariant toric divisor in $\widetilde{Y}$ via the pull-back of the projection: $\widetilde{\pi}: \widetilde{Y} \to \widetilde{X}$.
$Q$-invariant Weil divisors of $\widetilde{Y}$ are generated by the following elements:
\begin{equation}
\sum_{g^\prime \in Q} \widetilde{D}_{[g^\prime], g^\prime g_1,   g^\prime g_2}
\end{equation}
where $g_1, g_2 = \pm I, \pm J, \pm K$ and $g_1 \neq -g_2$. The number of such elements is 9.
Hence, 
\begin{equation}
\mathrm{WDiv}_T(\widetilde{X}) \simeq \BZ^9.
\end{equation}
The linear equivalence on $\mathrm{WDiv}_T(\widetilde{X})$ is trivial. Linearly equivalent divisors will induce the linearly equivalent divisors by the pull-back, $\widetilde{\pi}^*$.  Note
\begin{equation}
\sum_{g^\prime \in Q} \xi_{[g^\prime], g^\prime g_1, g^\prime g_2} = 0
\end{equation}
for any $g_1, g_2 \in Q$. This implies that no non-zero $Q$-invariant toric divisor in $\widetilde{Y}$ is linearly trivial and hence, 
\begin{equation}
T\mathcal{K}_{\widetilde{X}} \simeq \BZ^9 \otimes \BC \simeq \BC^9.
\end{equation}

We conclude this section by mentioning that $Q$-action on $\widetilde{V}$ will induce the original $Q$-action on $\BP^7$ via the monomial-divisor mirror map between the complex structure moduli space of $\widetilde{Y}$ and the K\"ahler moduli space of $Y$. The lattice points of $\nabla_{[g]}$ have the dual role as before: representing monomials in $\widetilde{G}_{[g]}$ and divisors in $\BP^7$. Each $\nabla_{[g]}$ has three such points: $0, e_g, e_{-g}$. The only common point $0$ represents 4 monomials, one from each $\widetilde{G}_{[g]}$. $Q$-action on the homogeneous coordinates $\widetilde{X}_{[g_1], g_2, g_3}$ of $\widetilde{V}$ will permute these 4 monomials.
Lattice points representing other monomials in $\widetilde{G}_{[g]}$ will be mapped to each other by $Q$-action on $\widetilde{X}_{[g_1], g_2, g_3}$ giving the original $Q$-action  on $Y$.

\section{Application}
$Q$ has a 2-dimensional irreducible representation, $V_2$, out of which one can build a flat rank-2 vector bundle on $X$.
In   \cite{Brunner:2001sk}, it was conjectured that there exists a threshold bound state of 2 D6-branes corresponding to this vector bundle  and this state becomes massless around the ``conifold'' point in the K\"ahler moduli space of $X$. In addition to this state, there are 4 more states that become massless. They are the line bundles associated to 1-dimensional irreducible representations, $V_1, V_I, V_J$, and $V_K$.
Unlike these line bundles, the existence of the flat rank-2 vector bundle as a stable single-particle state was not guaranteed by the BPS condition (it is degenerate with a pair of D6-branes), nor by K-theory (it does not carry any K-theory charge by which it might be distinguished from a pair of D6-branes). 
It is hard to check the existence directly on $X$ since we are far from the large radius limit.
Mirror symmetry gives a way of confirming it because the classical consideration on $\widetilde{X}$ is exact.

First, let's consider the complex structure moduli space of $\widetilde{X}$ and find the conifold point. Since it is independent of where we are in the K\"ahler moduli space, we can work on $\widehat{X} = \widehat{Y} / Q$, instead. It turns out that $\psi^8$ is the invariant parameter. To see this consider the following linear transformation:
\begin{equation}
\widehat{X}_{[g], 1} \mapsto \zeta \widehat{X}_{[g], 1} 
\end{equation}
with $\zeta^8 = 1$ and other homogeneous coordinates fixed. This transformation commutes with the $Q$-action modulo gauge transformation (\ref{GaugeXtilde}). This will change $\psi$ to $\zeta \psi$.
It is not difficult to show that $\psi^8$ is invariant under linear transformations that keep $\widehat{G}_{[g]}$'s in the form (\ref{Ghat}). This is consistent with the discussion we had earlier. $\widehat{Y}$'s invariant parameter is also $\psi^8$. Therefore, the moduli space of $\widehat{X}$ is the same as that of $\widehat{Y}$.

It is obvious that $\widehat{X}$ becomes singular at $\psi = 0, \infty$. They are the hybrid point and the large radius limit of $X$. There is another value of $\psi^8$ where $\widehat{X}$ becomes singular.
At singular points on the complex moduli space of $\widehat{X}$, there exist non-trivial solutions $P_{[g]}$ to the following equation:
\begin{equation}
\sum_{[g] \in Q/[Q,Q]} P_{[g]} \mathrm{d} \widehat{G}_{[g]} = 0.
\end{equation}
By non-triviality, we mean that not every $P_{[g]}$ is zero. With assumption $\psi \neq 0, \infty$, one can show that the previous equation implies
\begin{equation}
\prod_{[g^\prime] \in Q/[Q,Q]} \widehat{X}_{[g^\prime], g}^2 = \prod_{[g^\prime] \in Q/[Q,Q]} \widehat{X}_{[-g^\prime], -g}^2.
\end{equation}
Recall that this was the $Q$-fixed point condition in (\ref{FixedCond}). Therefore, the transversality fails at $Q$-fixed points and it happens only when $\psi^8 = 1$. The detailed calculation shows that there is only one such point in $\widehat{X}$ and it is given by $\widehat{X}_{[g], g^\prime}  = 1$ when $\psi = 1$.
This is the conifold point of $\widehat{X}$. 
To see how many states become massless, we expand around this point. Since non of the homogeneous coordinates is zero, we can make the following gauge choice:
\begin{equation}\label{GaugeChoice}
\widehat{X}_{[1],1} = \widehat{X}_{[g], g^\prime} = 1
\end{equation}
where $[g] \neq [1]$. Basically, we gauged away all homogeneous coordinates but $\widehat{X}_{[1], g}, g \neq 1$. Around the conifold point, we set 
\begin{gather}
\psi = 1 + \frac{\epsilon}{8}\notag\\
\widehat{X}_{[1], -1} = 1+y_1 \\
\widehat{X}_{[1], I} = \sqrt{\psi} (1 + \frac{y_2}{2})\;,\; \ldots\;,\; \widehat{X}_{[1], -K}=\sqrt{\psi} (1 + \frac{y_7}{2})\notag.
\end{gather}
 Assuming $\epsilon \ll 1$ and $y_i \sim \mathcal{O}(\sqrt{\epsilon})$, we get the following from $\widehat{G_1}, \ldots, \widehat{G_4}$ (up to the first order in $\epsilon$).
\begin{gather}
\epsilon  = y_1^2 + y_3^2 + y_5^2 + y_7^2 \notag\\
y_2  = -y_3 - y_3^2\notag\\
y_4  = -y_5- y_5^2\\
y_6 = -y_7-y_7^2\notag
\end{gather}
This is the cotangent space of $S^3$. If $\epsilon$ is real positive, then $S^3$ is parameterized by the real part of $y_1, y_3$ and $y_5$. Now, let's consider how $Q$ acts on it. $Q$-action must be followed by appropriate gauge transformations to maintain the gauge choice (\ref{GaugeChoice}). Since $I$ and $J$ generate the entire $Q$, it is enough to describe their action:
\begin{equation}
\begin{split}
I &: y_1 \mapsto -y_3,\; y_3 \mapsto y_1,\; y_5 \mapsto - y_7,\; y_7 \mapsto y_5\\
J &: y_1 \mapsto -y_5,\; y_3 \mapsto y_7\;, y_5 \mapsto y_1,\; y_7 \mapsto -y_3
\end{split}
\end{equation}
This representation is isomorphic to $V_2 \oplus V_2$ and is real.  $Q$ maps $S^3$ into itself. Also, $Q$ acts freely on $S^3$, since there is no $Q$-fixed point. Hence, the actual 3 cycle that shrinks to size zero is $S^3/Q$. Supersymmetric D3-branes wrapped on this 3 cycle have flat vector bundles. We can put 5 different vector bundles on $S^3/Q$; 4 from $V_1, V_I, V_J$ and $V_K$ and one from $V_2$. 
Therefore, there are 5 distinct states that become massless at the conifold point confirming the conjecture. 
Furthermore, from the mirror map, we know that the mirror of the flat rank-2 vector bundle on $S^3 / Q$ has D6-brane charge 2 and indeed it is the threshold bound state of 2 D6-branes.

Since we know which D-branes become massless at the conifold point, let's consider how the quantum symmetry group acts on them. Under tensor product, the flat line bundles form a group that is isomorphic to the quantum symmetry group. It is conjectured in   \cite{Brunner:2001eg}, on the level of K-theory, that the quantum symmetry group acts on the D-brane charges by tensoring the flat line bundles. Note that this conjecture applies to both A-branes and B-branes. Therefore, it will be a good test to see if the conjectured action is compatible with mirror symmetry.  On both $X$ and $\widehat{X}$, we have 4 flat line bundles built from the representations, $V_1, V_I, V_J$ and $V_K$. From the representation ring (\ref{IrrepsRing}) we considered earlier, it is clear that the 4 states built from one-dimensional irreducible representations form the regular representation of the quantum symmetry group and the state with D6-brane charge 2 is invariant. This action is compatible with mirror symmetry.

Unfortunately, the only known example of Calabi-Yau manifolds with non-abelian fundamental groups is the Beauville manifold. With more examples, we can check more throughly the conjectures we considered here. Hence, finding more examples of such 
Calabi-Yau manifolds is desirable.  
\section*{Acknowledgements}
I would like to thank P.~Candelas for helpful conversations and C.~Krishnan for proof-reading. This paper originated in and evolved through dicusssions with J.~Distler. I am deeply grateful to him. This research was supported in part by the National Science Foundation under Grant No. PHY0071512.




\bibliography{Beauville}
\bibliographystyle{utphys}

\end{document}